\documentclass{PoS}
\usepackage[numbers]{natbib}
\usepackage{float}
\usepackage{wrapfig}
\usepackage{graphicx}
\usepackage{amssymb}
\usepackage{bm}
\usepackage{amsmath}
\usepackage{physics}
\usepackage{amsthm}
\usepackage{cancel} %NEW
\usepackage{tabularx} %NEW
\usepackage{siunitx} %NEW
\usepackage{comment}
\usepackage{multicol}
\usepackage[]{caption}
\mathchardef\mhyphen="2D
\usepackage[hyphens,spaces,obeyspaces]{url}
\title{Estimation of CP violating EDMs from known mechanisms in the SM
\thanks{This work was supported by SERI-FCS award \# 2015.0594, Sigma Xi grants G2017100190747806 and G2019100190747806, and DOE grant DE-SC0014448. We would like to thank Y. Stadnik, N. Yamanaka, and K. S. Kirch for useful discussions.}}

\ShortTitle{Estimation of EDMs in the SM-$\{CKM\bigoplus\bar{\theta}\}$ framework}

\author{\speaker{Prajwal Mohanmurthy}\\
        Laboratory for Nuclear Science, Massachusetts Institute of Technology\\ 77 Mass. Ave., Bldg 26-540, Cambridge, MA 02139\\
        E-mail: \email{prajwal@mohanmurthy.com}}

\author{Jeff A. Winger\\
        Department of Physics and Astronomy, Mississippi State University\\
        PO Box 5167, Mississippi State, MS 39752\\
        E-mail: \email{j.a.winger@msstate.edu}}

\abstract{
New sources of CP violation, beyond the known sources in the standard model (SM), are required to explain the baryon asymmetry of the universe. Measurement of a non-zero permanent electric dipole moment (EDM) in fundamental particles, such as in an electron or a neutron, or in nuclei or atoms, can help us gain a handle on the sources of CP violation, both in the SM and beyond. Multiple mechanisms within the SM can generate CP violating EDMs, \emph{viz.} through the CKM matrix in the weak sector or through the QCD $\bar{\theta}$ parameter in the strong sector. We will estimate the maximum possible EDMs of leptons, certain baryons, select atoms and molecules in the (CKM$\bigoplus\bar{\theta}$) framework, assuming that the EDM wholly originates from either of the two SM mechanisms, independently. These estimates have been presented in light of the current experimental upper limits on the EDMs, in the following systems - leptons: $e^-$, $\mu^-$, $\tau^-$, $\nu_e^0$, $\nu^0_{\mu}$, $\nu^0_{\tau}$, baryons: $n^0$, $p^+$, $\Lambda^0$, $\Sigma^0$, $\Xi^0$, $\Lambda^+_c$, $\Xi^+_c$, atoms: $^{85}$Rb, $^{133}$Cs, $^{210}$Fr, $^{205}$Tl, $^{199}$Hg, $^{129}$Xe, $^{225}$Ra, $^{223}$Rn, and molecules: HfF$^+$, PbO, YbF, ThO, RaF, TlF. EDMs in different systems constrain CP-violating interactions differently \emph{i.e.} the same measured constraint on the EDM in two different systems may not actually be equally constraining on CP violating parameters. Finally, we emphasize the need to measure a non-zero EDM in multiple systems before understanding the origins of these CP-violating EDMs.
}
\FullConference{The 40th International Conference on High Energy Physics (ICHEP 2020)\\
		28 July - 6 August 2020.\\
		Prague, Czech Republic}

\begin{document}

\section{Introduction}

A non-zero electric dipole moment (EDM) in sub-atomic particles indicates violation of parity (P) and time reversal (T) symmetries. The standard model (SM) conserves the joint symmetry of charge inversion, parity, and time reversal (CPT) \cite{Luders1957-cy}, and consequently a non-zero EDM is CP violating. CP violation is a required condition for baryogenesis in the early universe \cite{Sakharov1967-vt}. The amount of CP violation is thus an important measurable.

CP violation is a key ingredient of weak interactions, and in the quark sector of the SM it is encoded in the Cabbibo-Kobayashi-Maskawa (CKM) quark oscillation matrix \cite{Cabibbo1963-kf,Kobayashi1973-tx}. Note that since the mass differences of charged leptons is large, the oscillation between charged leptons is suppressed, but there is also CP violation encoded into the Pontecorvo-Maki-Nakagawa-Sakata (PMNS) matrix which describes the neutrino oscillations \cite{Pontecorvo1958-ik, Maki1962-yx}, but we do not consider the contributions from the PMNS matrix in this paper. The amount of CP violation arising from the CKM matrix in the SM in insufficient to explain the observed baryon asymmetry of the universe \cite{Riotto1999-za}. CP violation could also be introduced into the strong sector of the SM, through the Quantum Chromodynamics (QCD)-$\bar{\theta}$ \cite{t_Hooft1976-lq}, but no CP violation has been observed in strong or electromagnetic interaction mediated processes.

In reality, the EDM of a system may arise from many sources: (i) the intrinsic EDM of the fundamental particles for leptons and quarks, $d_i$, (ii) CP violating pion-nucleon or nucleon-nucleon (both long range $\pi$NN-isoscalar, $\bar{g}^{(0)}_{\pi}$, isovector, $\bar{g}^{(1)}_{\pi}$, isotensor, $\bar{g}^{(2)}_{\pi}$, and short range hadronic interactions, $\bar{d}^{(SR)}_{h}$), and (iii) CP violating electron-nuclear (both scalar, $C_S$, and tensor, $C_T$) interactions. In a low energy paradigm, the EDM of a species, $d$, maybe written as \cite{Engel2013-rk,Chupp2015-ns,Chupp2019-cm}:
\begin{equation}
d = \sum_{i}\kappa_{d_i} d_i + \sum_{h\in\{n,p\}} \kappa_{\bar{d}^{(SR)}_{h}} \bar{d}^{(SR)}_{h} + \sum_{j\in\{0,1,2\}} \kappa_{\bar{g}^{(j)}_{\pi}} \bar{g}^{(j)}_{\pi} + \kappa_{C_S} C_S + \kappa_{C_T} C_T, \label{eqEDM-1}
\end{equation}
where $i  \in \{ u,d,s,c,t,b,e,\mu,\tau,\nu_e,\nu_{\mu},\nu_{\tau} \}$, and $\kappa_k$ are coefficients showing the sensitivity of the EDM to the particular species \emph{w.r.t.} the corresponding source. Measurement of a statistically significant EDM in multiple species helps us better understand the sources of CP violation. In this paper we will focus on reviewing and estimating the EDMs generated by a single source, coming from either of two mechanisms in the SM, CKM or QCD-$\bar{\theta}$, and place them in light of the current experimental constraints. We have only considered systems where experimental constraints already exist or have been planned. Furthermore, we have neglected the sign of the EDM. In the following sections we will describe the process for estimation of EDMs in subatomic particles, in atoms, as well as in molecules. Finally, we have concluded with a short note discussing the short comings of these estimates. A detailed review of the various effects contributing to the EDMs of these systems may be found in Refs. \cite{Engel2013-rk,Chupp2015-ns,Chupp2019-cm}.

\section{EDM of sub-atomic particles}

\paragraph*{{\bf ${\bm e^-}$:}} Diagrams that contribute to the EDM of charged leptons have four-loops \cite{Pospelov2014-gx}. This suppresses the SM-CKM mechanism generated electron EDM to $d^{\text{\tiny{(SM-CKM)}}}_e\sim10^{-44}~$e$\cdot$cm \cite{Pospelov2014-gx}. Ref. \cite{Ghosh2018-ob} has shown that electrons may also acquire an EDM through the SM-$\bar{\theta}$ of about $d^{\text{\tiny{(SM-$\bar{\theta}$)}}}_e\sim8.6\times10^{-38}~$e$\cdot$cm. Using the polar molecule ThO, the ACME collaboration has set an experimental constraint of $d^{\text{\tiny{(90\% C.L.)}}}_e < 1.1\times10^{-29}~$e$\cdot$cm \cite{ACME_Collaboration2018-ny}.

\paragraph*{{\bf $\bm \mu^-(\bm \tau^-)$:}} The mechanisms that generate an EDM in muons (tau leptons) are identical to those in electrons. The EDM of charged leptons acquired through the two SM mechanisms scales with the inverse ratio of the masses \cite{Ghosh2018-ob}. Thus, we can scale up the SM-CKM and the SM-$\bar{\theta}$ EDMs of the electron by a factor of $m_{\mu^-(\tau^-)}/m_e$ to obtain the theoretical SM expectation for the muon (tau lepton) EDM. The experimental constraint for the muon (tau lepton) EDM, $d^{\text{\tiny{(90\% C.L.)}}}_{\mu^-(\tau^-)} < 0.16(38)\times10^{-18}~$e$\cdot$cm, comes directly from the muon $g\mhyphen 2$ experiment \cite{Muon_g-2_Collaboration2009-qh} (comes from the Belle collaboration, using $e^-e^+$ collisions \cite{Inami2003-sv}). More recently, Ref.~\cite{Kirch2020-dr} has updated the limit to $d^{\text{\tiny{(90\% C.L.)}}}_{\tau^-} < 2.6\times10^{-18}~$e$\cdot$cm using the above method.

\paragraph*{{\bf ${\bm \nu^0}$:}} \sloppy The theoretical expectations for EDMs of neutrinos is as of yet unavailable. All the experimental constraints for neutrino EDMs come from indirect measurements involving $e\bar{e}$ collisions. The constraints of $d^{\text{\tiny{(90\% C.L.)}}}_{\nu_e^0} < 2\times10^{-21}~$e$\cdot$cm and $d^{\text{\tiny{(90\% C.L.)}}}_{\nu^0_{\mu}} < 2\times10^{-21}~$e$\cdot$cm have been reported in Refs. \cite{Del_Aguila1990-gg,Okun1986-oj}. Refs. \cite{Ibrahim2010-lr,Gutierrez-Rodriguez2004-xs,Escribano1997-lz} have reported a constraint of $d^{\text{\tiny{(90\% C.L.)}}}_{\nu_{\tau}} < 4.35\times10^{-17}~$e$\cdot$cm.

\paragraph*{n$^0$, p$^+$:} Diagrams that generate an EDM in baryons are simpler $\pi NN$ diagrams with just one-loop \cite{Khriplovich1982-jh}, and consequently we can expect larger baryonic SM EDMs when compared to charged leptons. The neutron's SM-CKM EDM has been estimated to be around $d^{\text{\tiny{(SM-CKM)}}}_{n}\sim 2\times10^{-32}~$e$\cdot$cm \cite{Khriplovich1982-jh}. The QCD-$\bar{\theta}$ parameter, in fact, is directly proportional to the neutron EDM value \cite{Pospelov2005-xn}:
\noindent
\begin{tabularx}{\textwidth}{XX}
\vspace{-0mm}
\begin{equation}
d_n \sim \bar{\theta} \cdot (6\times10^{-17})~\text{e$\cdot$cm}, \label{eq3}
\end{equation} &
\vspace{-0mm}
\begin{equation}
\bar{\theta}^{\text{\tiny{(90\% C.L.)}}}<3\times10^{-10}. \label{eq4}
\end{equation}
\end{tabularx}
Therefore, the maximum allowed contribution of the QCD-$\bar{\theta}$ parameter is constrained by the experimental upper limit on the neutron EDM, $d^{\text{\tiny{(90\% C.L.)}}}_{n} < 1.8\times10^{-26}~$e$\cdot$cm \cite{Abel2020-jr}, which implies the constraint in Eq.~\ref{eq4}, making the maximum allowed value of $d^{\text{\tiny{(SM-$\bar{\theta}$)}}}_{n}$ the same as well. Therefore, we do not need any physics beyond the standard model (BSM) effects to generate a neutron EDM. Any neutron EDM measured above the SM-CKM value would just be attributable to the QCD-$\bar{\theta}$ parameter. However, QCD-$\bar{\theta}$ could be close to zero, in which case the neutron could have an EDM wholly generated by SM-CKM, or in addition to possible BSM sources. If the neutron EDM was solely responsible for the $^{199}$Hg EDM, then the constraint extracted from the measurement of the $^{199}$Hg EDM is in fact better than the direct neutron EDM measurement using ultra-cold neutrons, $d^{\text{\tiny{$^{199}$Hg, (90\% C.L.)}}}_n<1.3\times10^{-26}~$e$\cdot$cm \cite{Graner2016-ge}.  However, we have adopted the direct measurement limit.

In $\pi NN$ diagrams (\emph{eg.} $p^+(n^0)\rightarrow n^0(p^+)\pi^{+(-)}$), one can easily interchange the proton with the neutron (and vice versa). This indicates that the SM-CKM contribution to the proton EDM is similar to that of the neutron EDM. Similarly, we will limit the contribution of QCD-$\bar{\theta}$ to the proton EDM, with the value of SM-$\bar{\theta}$ EDM obtained from the neutron. If the proton EDM was solely responsible for the $^{199}$Hg EDM, then the constraint extracted from the measurement of the $^{199}$Hg EDM is $d^{\text{\tiny{(90\% C.L.)}}}_{p} < 1.7\times10^{-25}~$e$\cdot$cm \cite{Graner2016-ge}. Similar to the muon $g-2$ storage ring measurement which gave rise to a constraint on the muon EDM, there are plans to directly measure the proton EDM using a storage ring \cite{Anastassopoulos2016-ts}.

\paragraph*{{\bf $\bm \Lambda^0,\bm \Sigma^0,\bm \Xi^0,\bm \Lambda^+_c,\bm \Xi^+_c$:}} The rest masses of the strange and charmed baryons, which we have considered, are within a factor of $2-3$ of each other, unlike the orders of magnitude variation of masses in charged leptons, so we neglect any mass effects here. Ref. \cite{Pich1991-zb} estimates that $d^{\text{\tiny{(SM-CKM)}}}_{\Lambda^0} \sim d^{\text{\tiny{(SM-CKM)}}}_{n}/2$, and Ref. \cite{Atwood1992-dd} shows the ratio between EDMs of the $\Lambda^0,~\Sigma^0,\text{and}~\Xi^0$ baryons; $\{d^{\text{\tiny{(SM-CKM)}}}_{\Lambda^0}:d^{\text{\tiny{(SM-CKM)}}}_{\Sigma^0}:d^{\text{\tiny{(SM-CKM)}}}_{\Xi^0}\}=\{3:-1:4\}$. Baryon EDM generating $\pi NN$ diagrams involve conversion of $\{u,s,b\}\leftrightarrow\{d,c,t\}$. The SM-CKM EDM for neutrons and protons are comparable, since their quark content involves exchanging a $u$ quark with a $d$ quark. Similarly, we obtain $\{\Lambda_c^+,\Xi_c^+\}$ baryons by exchanging the $s$ quark of $\{\Lambda^0,\Xi^0\}$ with a $c$ quark, respectively. Consequently, we estimated the SM-CKM EDMs of the two charmed baryons as $d^{\text{\tiny{(SM-CKM)}}}_{\Lambda^0} \sim d^{\text{\tiny{(SM-CKM)}}}_{\Lambda_c^+}$ and $d^{\text{\tiny{(SM-CKM)}}}_{\Xi^0}\sim d^{\text{\tiny{(SM-CKM)}}}_{\Xi_c^+}$. Refs. \cite{Borasoy2000-lp,Guo2012-fl}, give the EDM arising from the QCD-$\bar{\theta}$ parameter for these baryons. But, coupled with the constraint upon $\bar{\theta}$ from the neutron EDM, the constraint on the SM-$\bar{\theta}$ EDM portion for these baryons is weaker than that for the neutron. Therefore, we will constrain the SM-$\bar{\theta}$ EDM for these baryons with the same constraint we used for the neutron or proton. For the strange and charmed baryons, the EDM of only one has been measured: $d^{\text{\tiny{(90\% C.L.)}}}_{\Lambda^0} < 9.1\times10^{-17}~$e$\cdot$cm \cite{Pondrom1981-qi} using data from $p^+ X$ collisions in a fixed target experiment.

\begin{table}[h]
\centering
\begin{tabular}{rlrrrl}
\hline
\hline
& & $|d^{\text{\tiny{(SM-CKM)}}}|<$ & $|d^{\text{\tiny{(SM-$\bar{\theta}$)}}}|<$ & $|d^{\text{\tiny{(90\% C.L.)}}}|<$ & Ref.\\
\hline
\textbf{Charged leptons: }&$e^-$ & $10^{-44}$ & $8.6\times10^{-38}$ & $1.1\times10^{-29}$ & \cite{ACME_Collaboration2018-ny}\\
&$\mu^-$ & $2.1\times10^{-42}$ & $1.8\times10^{-35}$ & $1.6\times10^{-19}$ & \cite{Muon_g-2_Collaboration2009-qh}\\
&$\tau^-$ & $3.6\times10^{-41}$ & $3.1\times10^{-34}$ & $2.6\times10^{-18}$ & \cite{Kirch2020-dr}\\
\hline
\textbf{Neutrinos: }&$\nu_e^0$ & $-$ & $-$ & $2\times10^{-21}$ &  \cite{Del_Aguila1990-gg}\\
&$\nu^0_{\mu}$ & $-$ & $-$ & $2\times10^{-21}$ &  \cite{Del_Aguila1990-gg}\\
&$\nu^0_{\tau}$ & $-$ & $-$ & $4.4\times10^{-17}$ &  \cite{Gutierrez-Rodriguez2004-xs,Escribano1997-lz}\\
\hline
\textbf{Light baryons: }&$n^0_{(udd)}$ & $2\times10^{-32}$ & $1.8\times10^{-26}$ & $1.8\times10^{-26}$ &  \cite{Abel2020-jr}\\
&$p^+_{(uud)}$ & $2\times10^{-32}$ & $1.8\times10^{-26}$ & $1.7\times10^{-25}$ &  \cite{Graner2016-ge}\\
\hline
\textbf{Strange \& charmed baryons: }&$\Lambda^0_{(uds)}$ & $1\times10^{-32}$ & $1.8\times10^{-26}$ & $9.1\times10^{-17}$ &  \cite{Pondrom1981-qi}\\
&$\Sigma^0_{(uss)}$ & $3.3\times10^{-33}$ & $1.8\times10^{-26}$ & $-$ &\\
&$\Xi^0_{(uss)}$ & $1.3\times10^{-32}$ & $1.8\times10^{-26}$ & $-$ &\\
&$\Lambda^+_{c(udc)}$ & $1\times10^{-32}$ & $1.8\times10^{-26}$ & $-$ &\\
&$\Xi^+_{c(usc)}$ & $1.3\times10^{-32}$ & $1.8\times10^{-26}$ & $-$ &\\
\hline
\hline
\end{tabular}
\caption{Table showing the reach of SM-CKM and SM-$\bar{\theta}$ contributions to the species' EDM. The numbers are in the units of e$\cdot$cm. }
\label{tab3}
\end{table}

\section{EDM of Atoms}

Atoms may acquire a permanent EDM owing to the EDM of the nucleus, electrons, or from the CP-violating interactions between the electrons and the nucleus (or other electrons). Ideally, the electron cloud shields any nuclear EDM effectively suppressing the contribution of the nuclear EDM to the atomic EDM \cite{Schiff1963-bz}. Schiff shielding is not perfect in cases where \cite{Liu2007-wk}: (i) the electrons are relativistic \cite{Bernreuther1991-xl}, especially in high-$Z$ paramagnetic atoms with a single unpaired electron like in the alkali atoms $^{85}$Rb, $^{133}$Cs and $^{210}$Fr, but also in $^{205}$Tl, and/or (ii) the nucleus has quadrupole and octupole deformations, like in diamagnetic atoms of $^{225}$Ra and $^{223}$Rn, and possibly also in $^{199}$Hg and $^{129}$Xe, and/or (iii) there exists dominant CP-violating interactions between constituents of the atoms. The semi-leptonic scalar \cite{ACME_Collaboration2018-ny} and tensor \cite{Graner2016-ge} e$^-$N interaction parameters have already been constrained, respectively:
\noindent
\begin{tabularx}{\textwidth}{XX}
\vspace{-8mm}
\begin{equation}
\vspace{-8mm}
C^{\text{\tiny{(90\% C.L.)}}}_S<7.2\times10^{-10}, \label{eq1}
\end{equation} &
\vspace{-8mm}
\begin{equation}
\vspace{-8mm}
C^{\text{\tiny{(90\% C.L.)}}}_T<1.3\times10^{-10}, \label{eq2}
\end{equation}
\end{tabularx}
where the paramagnetic atoms (and polar molecules) are sensitive to the $C_S$ parameter and the diamagnetic atoms (and molecules) are sensitive to the $C_T$ parameter.

\paragraph*{$^{85}$Rb, $^{133}$Cs, $^{210}$Fr, $^{205}$Tl:} An enhanced electron EDM along with the semi-leptonic scalar e$^-$N interactions are responsible for the atomic EDM in these paramagnetic atoms: $d_{\text{\tiny{Atom}}}\sim(\partial d_{\text{\tiny{Atom}}}/\partial d_e)d_e+(\partial d_{\text{\tiny{Atom}}}/\partial C_S) C_S$. In order to estimate the SM-CKM EDM for these atoms, the SM-CKM EDM of the electron and the constraint upon $C_S$, in Eq.~\ref{eq1}, have been scaled up by their respective factors in Table~\ref{tab1}, and added together. For paramagnetic atoms, additional contributions to $C_S$ enter through $\bar{\theta}$ \cite{Flambaum1985-wv}, and after combining with the constraint upon $\bar{\theta}$ in Eq.~\ref{eq4} yields:
\begin{equation}
C_S(\bar{\theta})\sim0.03\bar{\theta}\implies C_S(\bar{\theta})<9\times10^{-12}.\label{eq5}
\end{equation}
The SM-$\bar{\theta}$ EDM values were obtained by scaling up the SM-$\bar{\theta}$ EDM of the electron and constraint on $C_S(\bar{\theta})$ above with their respective scaling factors in Table~\ref{tab1}, and added together.

\begin{table}[h]
\centering
\begin{tabular}{r|l|l}
\hline
\hline
Atom & |$\partial d_{\text{\tiny{Atom}}}/\partial d_e$| & |$\partial d_{\text{\tiny{Atom}}}/\partial C_S$|\\
\hline
$^{89}$Rb & $~~25.7$ \cite{Nataraj2008-tj}& $~~1.2$ \cite{Nataraj2008-tj}\\
$^{210}$Fr & $903.~~$ \cite{Byrnes1999-rv}& $~~5.0$ \cite{Chupp2015-ns}\\
$^{133}$Cs & $123.~~$ \cite{Nataraj2008-tj}& $~~7.1$ \cite{Chupp2019-cm}\\
$^{205}$Tl & $573.~~$ \cite{Liu1992-pz}& $70.~~$ \cite{Chupp2019-cm}\\
\hline
\hline
\end{tabular}
\caption{Sensitivity to atomic EDM \emph{w.r.t.}\ the e$^-$ EDM \cite{Engel2013-rk} and semi-leptonic e$^-$N interaction parameter, $C_S$ {\tiny ($\times10^{-19}~\text{e.cm}$)}, in the paramagnetic atoms.}
\label{tab1}
\end{table}

\paragraph*{$^{199}$Hg:} This nucleus may posses an intrinsic CP-violating nuclear Schiff moment which leads to a nuclear EDM \cite{Schiff1963-bz}. A non-zero nuclear Schiff moment can give rise to an amplified EDM in an atom depending on the electron configuration and its CP-violating interactions with the nuclear Schiff moment \cite{Flambaum1985-wv}. $^{199}$Hg and $^{129}$Xe are particularly interesting diamagnetic atoms whose atomic EDM is thought to have contributions from the EDMs of the nucleons, electrons, and most importantly from a non-zero nuclear Schiff moment. The atomic EDM of $^{199}$Hg \cite{Engel2013-rk} and its nuclear Schiff moment \cite{Dmitriev2003-im} can be decomposed to contributions from neutrons, protons, and electrons as:
\noindent
\begin{tabularx}{\textwidth}{XX}
\vspace{-10mm}
\small
\begin{equation}
\vspace{-7mm}
d_{\text{$^{199}$Hg}}=\rho_p\cdot d_p+\rho_n\cdot d_n+\kappa_S\cdot S + \mathcal{O}(d_e), \label{eq1-6-2-1}
\end{equation} &
\small
\vspace{-10mm}
\begin{equation}
\vspace{-7mm}
S_{\text{$^{199}$Hg}}=\left(0.20~\text{fm}^2\right)d_p+\left(1.895~\text{fm}^2\right)d_n,\label{eq1-6-2-2}
\end{equation}
\end{tabularx}
where $\kappa_S=-2.4\times10^{-4}~\text{fm}^2$ \cite{Graner2016-ge}, $\rho_p = (-0.56\times10^{-4})$ \cite{Engel2013-rk}, and $\rho_n = (-5.3\times10^{-4})$ \cite{Engel2013-rk}. Given that the electron SM-CKM EDM and its allowed SM-$\bar{\theta}$ EDM are about 12 orders of magnitude lower than that of the proton or neutron, we can safely neglect the contribution of the $d_e$ to the atomic EDM of $^{199}$Hg. Using the SM-CKM EDM and SM-$\bar{\theta}$ EDM values for the neutron and proton, we can write the theoretical expectations for the nuclear Schiff moment of $^{199}$Hg as: $|S^{\text{\tiny{(SM-CKM)}}}_{\text{\tiny{$^{199}$Hg}}}|\sim4.2\times10^{-19}~$e$\cdot$fm$^3$ and $|S^{\text{\tiny{(SM-$\bar{\theta}$)}}}_{\text{\tiny{$^{199}$Hg}}}|<6.3\times10^{-13}~$e$\cdot$fm$^3$. Along with the Schiff moment values calculated here, and the values of SM-CKM EDM and SM-$\bar{\theta}$ EDM associated with $d_p$ and $d_n$, we arrived at $\{|d^{\mathcal{T}\text{\tiny{(SM-CKM)}}}_{\text{\tiny{$^{199}$Hg}}}|,|d^{\mathcal{T}\text{\tiny{(SM-$\bar{\theta}$)}}}_{\text{\tiny{$^{199}$Hg}}}|\}=\{2.2\times10^{-35},6.2\times10^{-30}\}~\text{e}\cdot\text{cm}$.

The dependence of SM-CKM EDM for $^{199}$Hg on $C_T$ is: $\partial d_{\text{\tiny{$^{199}$Hg}}}/\partial C_T=3\times10^{-20}$ \cite{Chupp2015-ns}, and combining this with the constraint in Eq.~\ref{eq2} essentially yields the experimental limit. Similarly, the dependence on $C_S$: $\partial d_{\text{\tiny{$^{199}$Hg}}}/\partial C_S=-5.9\times10^{-22}$ \cite{Chupp2015-ns}, combined with the constraint in Eq.~\ref{eq1} gives $4.3\times10^{-31}~\text{e}\cdot\text{cm}$, while combining the above dependence \emph{w.r.t.} $C_S$ with Eq.~\ref{eq5}, yields $5.3\times10^{-33}~\text{e}\cdot\text{cm}$. On the other hand, $C_S(\bar{\theta})$ in Eq.~\ref{eq5} combined with the $\partial d_{\text{\tiny{$^{199}$Hg}}}/\partial C_S$ coefficient gives  $5.3\times10^{-33}~$e$\cdot$cm, so we shall neglect this and retain the SM-$\bar{\theta}$ previously obtained through the Schiff moment. In the case of $^{199}$Hg, any statistically significant EDM can be explained within the SM CKM through $C_S$ or $C_T$ parameters or the SM-$\bar{\theta}$ framework through the QCD-$\bar{\theta}$.

In this estimation, we've used $\kappa_S=-2.4\times10^{-4}~\text{fm}^2$ from Ref. \cite{Graner2016-ge}, but if we were to use  $\kappa_S=+2.8\times10^{-4}~\text{fm}^2$ from Ref. \cite{Engel2013-rk}, we'd obtain: $\{d^{\text{\tiny{(SM-CKM)}}}_{\text{\tiny{$^{199}$Hg}}}\sim 0,~d^{\text{\tiny{(SM-$\bar{\theta}$)}}}_{\text{\tiny{$^{199}$Hg}}}< 6\times10^{-32}\}~$e$\cdot$cm. It is important to note that the SM EDM of $^{199}$Hg could be zero. This implies that the constraint on $\bar{\theta}$ obtained from $^{199}$Hg is much more stringent than the value we extracted using neutron EDM ($\theta^{n}_s<5\times10^{-5}$). But due to the model dependence of the extraction of the constraint on QCD-$\bar{\theta}$ from $^{199}$Hg EDM, we will continue to rely on the value of QCD-$\bar{\theta}$ obtained from neutron EDM.

\paragraph*{$^{129}$Xe:} Scaling the values of $\{|d^{\mathcal{T}\text{\tiny{(SM-CKM)}}}_{\text{\tiny{$^{199}$Hg}}}|,|d^{\mathcal{T}\text{\tiny{(SM-$\bar{\theta}$)}}}_{\text{\tiny{$^{199}$Hg}}}|\}$ by the $\mathcal{T}$-factor indicated in Table~\ref{tab6} yields, $\{|d^{\mathcal{T}\text{\tiny{(SM-CKM)}}}_{\text{\tiny{$^{129}$Xe}}}|,|d^{\mathcal{T}\text{\tiny{(SM-$\bar{\theta}$)}}}_{\text{\tiny{$^{129}$Xe}}}|\}=\{3.0\times10^{-36},8.4\times10^{-31}\}~\text{e}\cdot\text{cm}$. By combining the limits in Eqs.~\ref{eq1} and \ref{eq2}, with $\partial d_{\text{\tiny{$^{129}$Xe}}}/\partial C_S=-4.4\times10^{-23}$ \cite{Chupp2015-ns} and $\partial d_{\text{\tiny{$^{129}$Xe}}}/\partial C_T$ from Table~\ref{tab6}, yields $3.2\times10^{-32}~\text{e}\cdot\text{cm}$ and $7.8\times10^{-31}~\text{e}\cdot\text{cm}$, respectively, clearly both larger than $d^{\mathcal{T}\text{\tiny{(SM-CKM)}}}_{\text{\tiny{$^{129}$Xe}}}$. Like in the case of $^{199}$Hg we will retain the value of $d^{\mathcal{T}\text{\tiny{(SM-$\bar{\theta}$)}}}_{\text{\tiny{$^{129}$Xe}}}$.

\paragraph*{$^{225}$Ra, $^{223}$Rn:} Higher electric and magnetic moments are not fully shielded by the Schiff screening. In such cases, not only is the nuclear Schiff moment contribution enhanced by the electron cloud, but there are additional enhancement factors contributing to the nuclear Schiff moment itself due to the octupole and quadrupole deformations of the nucleus \cite{Dzuba2002-or, Mohanmurthy2020-np}. Contributions from individual nucleons, or electrons here is negligible. Scaling up  $\{|d^{\mathcal{T}\text{\tiny{(SM-CKM)}}}_{\text{\tiny{$^{199}$Hg}}}|,|d^{\mathcal{T}\text{\tiny{(SM-$\bar{\theta}$)}}}_{\text{\tiny{$^{199}$Hg}}}|\}$ using the factors in Table~\ref{tab6} gives the Schiff moment contributions: $\{|d^{\mathcal{T}\text{\tiny{(SM-CKM)}}}_{\text{\tiny{$^{225}$Ra($^{223}$Rn)}}}|,|d^{\mathcal{T}\text{\tiny{(SM-$\bar{\theta}$)}}}_{\text{\tiny{$^{225}$Ra($^{223}$Rn)}}}|\}=\{1.6(0.63)\times10^{-32},4.5(1.8)\times10^{-27}\}~\text{e}\cdot\text{cm}$. Neglecting the contributions of $C_S$ (and any propagation of $C_S$ via the QCD-$\bar{\theta}$) like in the case of $^{129}$Xe, and combining the $\partial d_{\text{Atom}}/\partial C_T$ coefficients in Table~\ref{tab6} with the constraint in Eq.~\ref{eq2} gives $\{d^{\mathcal{T}\text{\tiny{(SM-CKM)}}}_{\text{\tiny{$^{225}$Ra($^{223}$Rn)}}}|=6.9(0.65)\times10^{-30}~\text{e}\cdot\text{cm}$.

\begin{table}[h]
\centering
\begin{tabular}{rrrrrl}
\hline
\hline
& & $|d^{\text{\tiny{(SM-CKM)}}}|<$ & $|d^{\text{\tiny{(SM-$\bar{\theta}$)}}}|<$ & $|d^{\text{\tiny{(90\% C.L.)}}}|<$ & Ref.\\
\hline
\textbf{Paramagnetic atoms:} & $^{85}$Rb & $8.6\times10^{-29}$ & $1.1\times10^{-30}$ & $1\times10^{-18}$ & \cite{Ensberg1967-kt}\\
& $^{133}$Cs & $5.1\times10^{-28}$ & $6.4\times10^{-30}$ & $1.1\times10^{-23}$ & \cite{Murthy1989-ay}\\
& $^{205}$Tl & $5.0\times10^{-27}$ & $6.3\times10^{-29}$ & $1.5\times10^{-24}$ & \cite{Commins1994-fe}\\
& $^{210}$Fr & $3.6\times10^{-28}$ & $4.5\times10^{-30}$ & $-$ & \\
\hline
\textbf{Diamagnetic atoms:} & $^{199}$Hg & $6.2\times10^{-30}$ & $6.2\times10^{-30}$ & $6.2\times10^{-30}$ & \cite{Graner2016-ge}\\
& $^{129}$Xe & $7.8\times10^{-31}$ & $8.4\times10^{-31}$ & $1.3\times10^{-27}$ & \cite{Allmendinger2019-el}\\
\hline
\textbf{Diamagnetic atoms with} & $^{225}$Ra & $6.9\times10^{-30}$ & $4.5\times10^{-27}$ & $1.2\times10^{-23}$ & \cite{Bishof2016-wu}\\
\textbf{nuclear octupole deformation: }& $^{223}$Rn & $6.5\times10^{-31}$ & $1.8\times10^{-27}$ & $-$ & \\
\hline
\hline
\end{tabular}
\caption{Table showing the expected SM-CKM and SM-$\bar{\theta}$ contributions to the species' atomic EDM, for $\{C_S, C_T\} \ne 0$. The numbers are in the units of e$\cdot$cm. }
\label{tab4}
\end{table}

\begin{table}[h]
\centering
\begin{tabular}{c|r|l|l}
\hline
\hline
Atom&\tiny{$\kappa^{\text{(Atom)}}_S/\kappa^{^{199}Hg}_S$}&\tiny{$d^{\mathcal{T}}_{\text{(Atom)}}/d^{\mathcal{T}}_{^{199}Hg}$}& \tiny{|$\partial d_{\text{\tiny{Atom}}}/\partial C_T$|}\\
\hline
${^{129}}$Xe&$1/7.4$ \cite{Dzuba2002-or}&$1$&$0.6$ \cite{Chupp2019-cm}\\
${^{225}}$Ra&$3.~~$\cite{Dzuba2002-or}&$240~$ \cite{Jesus2005-zv}&$5.3$ \cite{Chupp2019-cm}\\
${^{223}}$Rn&$~1.2$ \cite{Dzuba2002-or}&$240~$ \cite{Spevak1997-bg}&$0.5$ \cite{Chupp2015-ns}\\
\hline
\hline
\end{tabular}
\caption{By columns, enhancement factors \emph{w.r.t.}\ that in $^{199}$Hg (i) nuclear Schiff moment to the atomic EDM, and (ii) nuclear quadrupole and octupole deformation to the nuclear Schiff moment; dependence of atomic EDM on (iii) semi-leptonic tensor e$^-$N interaction {\tiny ($\times10^{-20}~$e$\cdot$cm)}.}
\label{tab6}
\end{table}

\section{EDM of Molecules}

All the experiments discussed so far, searching for EDMs in sub-atomic particles and atoms, applied an electric field, $E$ on the order of $\sim$kV/cm, to induce (Stark) energy level splitting owing to a possible non-zero EDM. Electric fields are subject to the break-down potential of vacuum. If large potentials are applied to two electrodes separated by a distance, then electrons begin to flow between the electrodes, limiting the effective potential difference at electric fields of about $E\sim30~k$V/cm. Recent, electric fields, as high as $E\sim112~k$V/cm, have been achieved using niobium electrodes \cite{BastaniNejad2012-xd}. Polar molecules contain charged atoms, which have large intra-molecular electric fields, $E_{\text{Mol.}}$, on the order of GV/cm. The EDM sensitivity achievable is directly proportional to the electric field we can apply. Experiments searching for EDMs in molecules use this intra-molecular electric field as a key to achieve higher EDM sensitivities compared to atomic EDM experiments.

\paragraph*{PbO, ThO, HfF$^+$, YbF:} In a polar molecule, the valence electrons of the atom feel the large intra-molecular electric field. Each of these polar molecules have a particular associated molecular electric field and the experiments measuring their EDM used an additional applied electric field, so their sensitivity to the electron EDM varies. In order to compare these molecular EDMs, we normalized the molecular EDM measured with the ratio of molecular electric field to the applied electric field, listed in Table~\ref{tab2}. The corresponding SM-CKM EDMs and SM-$\bar{\theta}$ EDMs are obtained using: $d^{\text{\tiny{\{(SM-CKM),(SM-$\bar{\theta}$)\}}}}_{\text{\tiny{Mol.}}}\sim[d^{\text{\tiny{\{(SM-CKM),(SM-$\bar{\theta}$)\}}}}_e+(\partial d_{\text{\tiny{Mol.}}}/\partial C_S)\{C^{\text{\tiny{(90\% C.L.)}}}_S,C_S(\bar{\theta})\}]\times(E_{\text{Mol.}}/E)$.
\begin{table}[h]
\centering
\begin{tabular}{l|l|c r|l}
\hline
\hline
Mol. & $E_{\text{Mol.}}$ \tiny{(GV/cm)}& $|d^{\text{\tiny{(90\% C.L.)}}}_e|<$ \tiny{(e$\cdot$cm)} & $E$  \tiny{(V/cm)} &|$\partial d_{\text{\tiny{Mol.}}}/\partial C_S$|\\
\hline
HfF$^+$ & $23~~~$ \cite{Fleig2013-dm} & $1.3\times10^{-28}$ & ${\bf 24}$  \cite{Cairncross2017-hc,Loh2013-jj} & $~~~~8.9$ \cite{Chupp2019-cm}\\
PbO & $25~~~$ \cite{Kozlov2002-tm,Petrov2005-tl} & $1.7\times10^{-26}$ & $\{{\bf 100},125\}$ \cite{Eckel2013-gv}~~~~~~~& $\sim4.2$ \cite{Flambaum2020-dz}\\
YbF & $14.5$ \cite{Kara2012-do} & $1.1\times10^{-27}$ & ${\bf 10k}~~~~~~$ \cite{Hudson2011-wy}~~~~~~~& $~~~~8.6$ \cite{Chupp2019-cm}\\
ThO & $78~~~$ \cite{Skripnikov2016-gh,Denis2016-zz} & $1.1\times10^{-29}$ & $\{{\bf 80},140\}$ \cite{ACME_Collaboration2018-ny}~~~~~~~& $~~13.~~$ \cite{Chupp2019-cm}\\
%\hline
%$^{225}$RaF & $0.13~$ & \cite{Skripnikov2016-gh,Denis2016-zz} & $1.1\times10^{-29}$ & $300~$ & \cite{ACME_Collaboration2018-ny}\\
\hline
\hline
\end{tabular}
\caption{Relevant intra-molecular electric field, $E_{\text{Mol.}}$, measured e-EDM, the external electric field, $E$, applied in the experiment, and the CP-odd semi-leptonic interaction parameter, $C_S$ {\tiny ($10^{-21}$e$\cdot$cm)}.}
\label{tab2}
\end{table}

\paragraph*{$^{225}$RaF:} The relevant molecular electric field felt by the radium nuclei in the $^{225}$RaF molecule is $E_{\text{\tiny{$^{225}$RaF}}}=130~$MV/cm \cite{Kudashov2014-ye}.
%In $^{225}$RaF, the relevant molecular electric field is that felt by the nucleus of $^{225}$Ra.
The nucleus of $^{225}$Ra is much heavier than an electron, so the electric field felt by the nucleus of $^{225}$Ra in its rest frame is much less than that in the polar molecules discussed above, where the same field is felt by the electron in its rest frame. The radium EDM experiment plans to apply an electric field of $E=300~$kV/cm \cite{BastaniNejad2012-xd} using niobium electrodes. The SM-CKM and SM-$\bar{\theta}$ contributions to the EDM of $^{225}$RaF were obtained by scaling up SM-CKM EDM and SM-$\bar{\theta}$ EDM values associated with atomic $^{225}$Ra with the ratio of $(E_{\text{Mol.}}/E)$.

\paragraph*{TlF:} The TlF molecule is a diamagnetic system, similar to the $^{199}$Hg atom. The EDM of TlF is $d_{TlF}\sim81d_e+d_p/2+2.9\times10^{-18}C_S+2.7\times10^{-16}C_T$ \cite{Chupp2019-cm,Cho1991-vt}. Clearly one could neglect the contribution of electron or proton EDMs. Hence, the SM-CKM EDM for TlF dominantly originates from $C_T$ with $|d^{\mathcal{T}\text{\tiny{(SM-CKM)}}}_{\text{\tiny{TlF}}}|=3.5\times10^{-26}~$e$\cdot$cm. SM-$\bar{\theta}$ contributions to the EDM of TlF arising from the proton is $9\times10^{-27}~$e$\cdot$cm, while the same originating from QCD-$\bar{\theta}$ and via the scalar semi-leptonic interaction can be neglected.

\begin{table}[h]
\centering
\begin{tabular}{rrrrl}
\hline
\hline
& $\frac{E_{\text{Mol.}}}{E}$ $\left(\times\frac{[(*)/\text{cm}]}{[(*)/\text{cm}]}\right)$ & $|d^{\text{\tiny{(SM-CKM)}}}|<$ & $|d^{\text{\tiny{(SM-$\bar{\theta}$)}}}|<$ & $|d^{\text{\tiny{(90\% C.L.)}}}|<$\\
\hline
\textbf{Diamagnetic mol.:} & TlF\hspace{9ex} & $3.5\times10^{-26}$ & $9\times10^{-29}$ & $4.8\times10^{-23}$ \cite{Cho1991-vt}\\
\hline
\textbf{Polar mol.:} & HfF$^+$ $\left(\frac{23~\text{GV}}{24~\text{V}}\right)$ & $6.1\times10^{-21}$ & $7.7\times10^{-23}$ & $1.2\times10^{-19}$ \cite{Cairncross2017-hc,Loh2013-jj}\\
& PbO~~~$\left(\frac{25~\text{GV}}{100~\text{V}}\right)$ & $7.6\times10^{-22}$ & $9.5\times10^{-24}$ & $4.3\times10^{-18}$ \cite{Eckel2013-gv}\\
& YbF $\left(\frac{14.5~\text{GV}}{10~\text{kV}}\right)$ & $9.0\times10^{-24}$ & $1.1\times10^{-25}$ & $1.5\times10^{-21}$ \cite{Hudson2011-wy}\\
& ThO~~ $\left(\frac{78~\text{GV}}{80~\text{V}}\right)$ & $1.1\times10^{-20}$ & $1.1\times10^{-22}$ & $1.1\times10^{-20}$ \cite{ACME_Collaboration2018-ny}\\
& RaF $\left(\frac{130~\text{MV}}{300~\text{kV}}\right)$ & $3.0\times10^{-27}$ & $2.0\times10^{-24}$ & $-$\\
\hline
\hline
\end{tabular}
\caption{Table showing the expected SM-CKM and SM-$\bar{\theta}$ contributions to the species' molecular EDM, for $\{C_S, C_T\} \ne 0$. The numbers are in the units of e$\cdot$cm. Here, $E_{\text{Mol.}}$ denotes the intra-molecular electric field experienced by the electron (HfF$^+$, PbO, YbF, ThO) or the nucleus (RaF), and $E$ denotes the applied electric field in the experiment.}
\label{tab5}
\end{table}

\begin{figure}[h]
\centering
    \includegraphics[width=\textwidth,angle=0,origin=c]{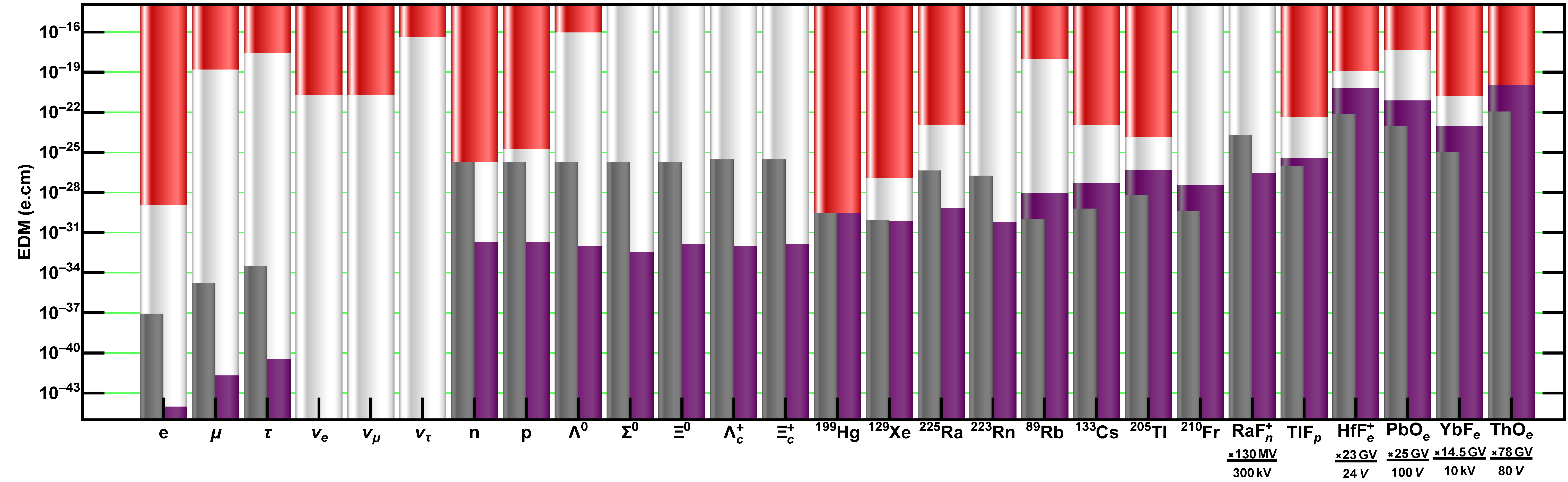}
\caption[]{Panel showing the measured upper limit and expected SM theoretical values of the EDM from the SM-CKM and SM-$\bar{\theta}$ mechanisms, for various species. The measured upper limit of EDM at 90\% C.L. has been shown in red. The gray portion represents the contribution of the QCD-$\theta_s$ parameter to the value of the EDM, and the purple portion shows the contribution of the SM-CKM matrix to the value of the EDM.}
\label{fig1}
\end{figure}

%%%%%%%%%%%%%%%%%%%%%%%%%%%%%%%%%%%%%%%%%%%%%%

\section{Conclusion}

As we go from the simplest to the more complex systems, \emph{i.e.} leptons to baryons to atoms to molecules, the possible mechanisms that can generate an EDM increases. Both the SM-CKM and SM-$\bar{\theta}$ portions of the EDM of paramagnetic atoms and polar molecules are dominated by the contributions of scalar semi-leptonic e$^-$N interactions, $C_S$, and the SM-CKM EDM of diamagnetic atoms are dominated by the tensor semi-leptonic e$^-$N interactions, $C_T$, while SM-$\bar{\theta}$ EDM is dominated by the Schiff moment contributions (mostly through long range $\pi$NN interactions). Furthermore, recent studies \cite{Yamaguchi2020-sr,Yamaguchi2020-gq} have also shown that the e$^-$-SM-CKM EDM could be much higher, using hadronic loops instead of quark 4-loops used in Ref. \cite{Pospelov2014-gx}, which may dramatically improve the estimates which depend on the e-EDM. These estimates of SM-CKM and SM-$\bar{\theta}$ EDMs for sub-atomic particles, have also been used in Refs. \cite{Kirch2020-dr,Alemany2019-cg,Beacham2019-zh}.

The constraints on $C_S$, $C_T$, and long range $\pi$NN interactions (through QCD-$\bar{\theta}$) come from the EDM measurements with ThO, $^{199}$Hg, and the neutron, respectively. Consequently any measurement of a statistically significant EDM in these three species could lead to a SM measurement, since mechanisms within the SM are sufficient to explain their EDM wholly. Note that  SM-CKM EDMs generally have a smaller reach than SM-$\bar{\theta}$ EDMs for all the fundamental particles, but the situation is reversed in paramagnetic atoms and polar molecules owing to constraints on QCD-$\bar{\theta}$ better constraining the EDM that it generates, compared to $C_S$ or $C_T$ (by $\sim2$ and $\sim3$ orders of magnitude, by comparing Tables~\ref{tab1} and \ref{tab6} with Eq.~\ref{eq3}, respectively). On the other hand, in the case of $^{199}$Hg, the reach of SM-CKM EDM is comparable to SM-$\bar{\theta}$ EDM, and clearly is a candidate where the constraint on QCD-$\bar{\theta}$ and $C_T$ comparably generate an EDM through the SM-$\bar{\theta}$ and SM-CKM mechanisms, respectively.

A statistically significant measurement in any one species would not help us understand the origins of its EDM, thereby requiring measurement of an EDM in multiple systems to parse out the contributions from various underlying mechanism. In Figure~\ref{fig1}, for sub-atomic particles, the white space between the SM-$\bar{\theta}$ EDM portion of the SM theoretical estimate and the experimental constraint could be fertile ground in which to search for physics BSM. Given that the white space for the charged leptons is the largest (over $\sim8$ orders of magnitude), they may be the apt systems in which to search for BSM effects given their low SM background. Additionally, new and improved efforts to measure the EDM of p$^+$, TlF, Ra, and RaF may make their status comparable to the current status of n$^0$, $^{199}$Hg, and ThO.

%\section*{References}
%The references maybe found in the full form of this paper at \href{www.arxiv.org/abs/}{arXiv:[]}
%\renewcommand{\bibpreamble}{The references maybe found in the full form of this paper at \href{www.arxiv.org/abs/}{arXiv:[]}}
%\begin{multicols}{2}
%\nocite{*}
%\bibliographystyle{aipnum-cp}%
\setlength{\bibsep}{1pt plus 1.0ex}

\end{document}